\documentclass[12pt,preprint]{aastex}

\newcommand{\uJy}{\ensuremath{\mbox{ $\mu$Jy}}}
\newcommand{\sgr}{SGR~1900+14}
\newcommand{\snr}{G42.8$+$0.6}
\newcommand{\sgrlmc}{\mbox{SGR~0526$-$66}}
\newcommand{\cxo}{\textit{CXO}}
\newcommand{\chandra}{\textit{Chandra}}
\newcommand{\xte}{\textit{RXTE}}
\newcommand{\rxte}{\textit{RXTE}}
\newcommand{\bepposax}{\textit{Beppo-SAX}}
\newcommand{\asca}{\textit{ASCA}}
\newcommand{\ctsec}{\mbox{c s$^{-1}$}}
\newcommand{\ctsecpcu}{\mbox{c s$^{-1}$ PCU$^{-1}$}}

\newcommand{\percm}{\mbox{cm$^{-2}$}}

\newcommand{\ergcms}{\mbox{erg cm$^{-2}$ s$^{-1}$}}
\newcommand{\ergcm}{\mbox{erg cm$^{-2}$}}
\newcommand{\expnt}[2]{\ensuremath{#1 \times 10^{#2}}}   % scientific notation
\def\mc {\multicolumn}

\begin{document}

\title{High-Resolution X-ray and Radio Observations of SGR~1900$+$14
in the Immediate Aftermath of a Giant Flare}

\author{D. W. Fox, D. L. Kaplan, S. R. Kulkarni}
\affil{Department of Astronomy, 105-24 California Institute of
Technology, Pasadena, California 91125, USA}
\email{derekfox@astro.caltech.edu, dlk@astro.caltech.edu,
srk@astro.caltech.edu} 

\and \author{D. A. Frail}
\affil{National Radio Astronomy Observatory, Socorro, NM 87801, USA}
\email{dfrail@nrao.edu}

\begin{abstract}
We present the results of \chandra, \rxte, and VLA observations of
\sgr\ in the immediate aftermath of its 2001 April~18 giant flare
event.  In the X-ray band we find the source in a pulsating and
bursting state, with time-averaged 2--10~keV flux initially elevated
by 20\% above the source's previous quiescent periods.  In the radio
we establish upper limits on the strength of any persistent post-flare
emission of 0.7 and 0.1~mJy at 1.4~GHz and 8.0~GHz, respectively.  The
position of the X-ray source is consistent, to approximately 1~arcsec
precision, with the August~1998 VLA determination, and the
one-dimensional X-ray profile is consistent with that of a point
source.  The X-ray spectrum is best-fit by a two component power-law
plus blackbody model, with fitted blackbody temperature $kT_{\rm
BB}\approx 0.5$~keV and radius $R_{\rm BB}\approx 1.5$~km for an
assumed distance of 5~kpc.  The spectral parameters of this thermal
component are consistent with those reported for the source in
quiescence, and the variations in the source flux we observe may be
explained as variations in the power-law component alone, providing
support for magnetar models of \sgr.  
\end{abstract}

\section{Introduction}
The Soft Gamma-Ray Repeaters (SGRs; see \citealt{h99} for a recent
observational review) are a unique class of Galactic neutron stars
that exhibit bright flaring activity in the hard X-ray and soft
gamma-ray bands.  Perhaps their most unusual and distinctive feature
is their hyper-Eddington bursts, the so-called ``giant flares,'' of
which the 5~March~1979 event \citep{mgi+79} is the most famous
example.  Based on the few observed events, the SGRs appear to emit
their giant flares on a recurrence timescale of years to decades.
They also exhibit softer quiescent X-ray emission with coherent
pulsations at periods of 5--10~s.

The SGRs are generally thought to be young ($<10^{4}$~yr) neutron
stars (NSs) with extremely strong magnetic fields ($>10^{14}$~G),
i.e.\ magnetars \citep{dt92,td93}.  This belief has been motivated by
their several associations with young supernova remnants (SNRs) or
star-forming regions \citep{ekl+80,kf93,fmc+99,vhl+00}, by the
energetics and phenomenology of their giant flares \citep{p92,td95},
and by the detection of X-ray pulsations with relatively long
(5--10~s) periods and large ($\sim 10^{-11}\mbox{ s s}^{-1}$)
spin-down rates \citep[e.g.][]{k+98,ksh+99}.

\subsection{\sgr}
Historically, multiple bursts from \sgr\ gave the first hint of the
existence of a new class of gamma-ray transients \citep{mgg79},
suggesting that \sgrlmc, source of the famous 5~March 1979 event, was
not alone.  Reactivation of \sgr\ in 1992 \citep{kfm+93} led to
follow-up observations that associated the burst source with a soft,
persistent X-ray source \citep{vkfg94,hlv+96} and, possibly, an SNR
12~arcmin distant \citep[\snr;][]{vkfg94}.  On 27~August 1998, a giant
flare with strong 5~s pulsations was detected by several spacecraft
\citep{hcm+99}; this flare remains the strongest gamma-ray event (in
peak flux and fluence) detected to date from any cosmic source apart
from the Sun.  Arrival-time localization \citep{hkw+99b} identified
\sgr\ as the source, and prompt radio observations detected a fading,
non-thermal radio source \citep{fkb99} coincident with the quiescent
X-ray source, a 5.17-s pulsar \citep{hlk+99}.  Timing analyses of the
quiescent X-ray emission soon revealed that the pulsar was spinning
down in magnetar-like fashion \citep{ksh+99}.  Observations of the
source in the aftermath of the flare demonstrated, in addition, a
short-term ($\sim$hours) increase in spin-down rate \citep{p01} and
associated deviations from simple spin-down \citep{wk+99}.  The radio
detection provided the highest-accuracy position for \sgr\ (current as
of August~1998): $\alpha_{2000}=19^{\rm h}07^{\rm m}14\fs33$,
$\delta_{2000}=+09\degr19\arcmin20\farcs1$, with uncertainty $\pm
0\farcs15$ in each coordinate \citep{fkb99}.

The precise localization of \sgr\ has since enabled further
observations at all wavelengths.  \citet{lx00} and \citet{k+01} have
raised questions about the association of \sgr\ with \snr;
\citet{vhl+00} have proposed instead an association of \sgr\ with an
embedded cluster of high-mass stars, 12~arcsec from the source, that
is probably located at a distance of 12--15~kpc.  Additional follow-up
observations in the infrared \citep{ed00}, optical, and radio
\citep{k+01b} have been undertaken; although these have not revealed
any point-source candidate counterparts to date, the upper limits on
persistent emission in the IR band can now constrain accretion disk
models for the source \citep[c.f.][]{vptvdh95,mlrh99}, as discussed by
\citet{k+00c}.

On 2001~April~18.33, the \bepposax\ and \textit{Ulysses} satellites
detected a second giant flare from \sgr\ \citep{g+01,h+01} with
duration $\sim 40$~s, 25-100~keV fluence $\sim \expnt{2.6}{-4}\mbox{
ergs cm}^{-2}$, and peak flux over 0.5~s of $\sim
\expnt{1.7}{-5}\mbox{ ergs cm}^{-2}\mbox{ s}^{-1}$ ($\sim 25\times$
less fluence and $\sim 200\times$ lower peak flux than the flare of
August~1998).  In response we initiated observations of \sgr\ with the
\textit{Chandra X-ray Observatory}, \textit{Rossi X-ray Timing
Explorer}, and Very Large Array, seeking to investigate the numerous
transient phenomena that have been associated with the giant flares.
\citet{ktw+01} have pursued a similar agenda with their overlapping
set of observations.

\section{Observations \& Analysis}
Our program sought to probe the immediate aftermath of the flare with
sensitive X-ray spectral, X-ray timing, and radio imaging
observations.  These observations also allow us to investigate the
nature of \sgr.

\subsection{Radio}
We observed the position of the radio transient associated with the
August~1998 giant flare \citep{fkb99} from \sgr\ with the Very Large
Array\footnote{The VLA is operated by the National Radio Astronomy
Observatory, which is a facility of the National Science Foundation
operated under cooperative agreement by Associated Universities, Inc.}
(VLA) in its ``B'' configuration on a number of occasions, as part of
our regular observing program and using observing time donated by
others.  A log of these observations is found in Table~\ref{tab:vla}.
The data were taken in continuum mode with 2$\times$50~MHz bandwidth.
They were reduced and calibrated using standard procedures in {\tt
AIPS}, and then imaged with the {\tt IMAGR} task.  This yielded beam
sizes of $\approx 4\farcs5$ and $0\farcs8$ at 1.4~GHz and 8.4~GHz,
respectively.  We did not detect a source in any of the observations,
giving the 3$\sigma$ upper limits plotted in Figure~\ref{fig:radio}.

\subsection{X-ray}
We observed \sgr\ with the {\em Chandra X-ray Observatory} on two
occasions after the flare, beginning on 2001~April~22.19 UT and
2001~April~30.97 UT, with durations of 20.8~ks and 18.9~ks
respectively, as part of the Director's Discretionary Time allocation,
with no proprietary data rights period.  Both observations were taken
in continuous-clocking mode with the aimpoint on the
backside-illuminated ACIS S-3 detector.  This gave us a time
resolution of 2.85~ms and mitigated the effects of photon pileup for
the persistent emission, as well as for bursts of modest size, while
sacrificing one dimension of spatial information.

The two \chandra\ observations were coordinated with two observations
of the \textit{Rossi X-ray Timing Explorer} (\xte) whose data were
also made immediately public.  Due to a change in \chandra\ planning
that was not mirrored by \xte, the first \chandra\ observation
occurred one day after the first \xte\ observation.  The second \xte\
and \chandra\ observations overlapped as planned.  The \xte\
observations began at 2001~Apr~21.32 UT and 2001~Apr~30.99 UT, lasted
for 15.2 and 15.6~ks respectively, and had total good-time intervals,
after screening of Earth occults and intervals of high electron
background, of 9.6 and 8.9~ks respectively.

\chandra\ data were processed, for the most part, according to
standard CXC procedures.  First we examined the lightcurve of a
background region for high-background intervals; none were identified.
We then extracted events recorded by the ACIS S-3 detector and
restricted to the energy range 0.3--10~keV.  We barycentered this data
using the {\tt axbary} tool with a preliminary \cxo\ ephemeris.  We
extracted the events from a region 10~pixels ($\approx 5\arcsec$)
wide. This region gives source count rates of $0.620 \pm 0.006\mbox{
s}^{-1}$ and $0.534\pm 0.005\mbox{ s}^{-1}$ for the two observations.
For precision timing analyses we were forced to account separately for
the charge-transfer time, that is, the approximately 4~s it takes for
the charge packets produced by each X-ray photon from the source to be
read out from the center of the ACIS-S3 chip.  We made this correction
in two different ways: first, by following an approximate prescription
related to us by the \chandra\ X-ray Center Helpdesk staff; and
second, by executing a shell script provided to us by Allyn Tennant of
the Marshall Space Flight Center\footnote{Script available at
\texttt{http://wwwastro.msfc.nasa.gov/xray/ACIS/cctime/}}.  Results of
the two approaches were identical; however, we note that the latter
approach is superior in that it incorporates higher-order corrections
for the dither-motion and flexure of the observatory over the course
of the observation; these corrections will make a difference for
analyses requiring significantly more precision than ours.

\xte\ realtime data were processed according to the protocols
described on the \xte\ web
site\footnote{\texttt{http://heasarc.gsfc.nasa.gov/docs/xte/recipes/cook\_book.html}};
note that the \xte\ pointing was offset from \sgr\ by 20~arcmin to
reduce contamination from the bright source GRS~1915+105, resulting in
a factor of 1.5 decrease in count rates relative to direct-pointing
observations due to reduced collimator efficiency.  The spectrum was
(particle) background-dominated at high energies, so timing analyses
were performed on 2--60~keV data only.

\subsubsection{Phase-Averaged Spectral Analysis}
Our spectral analysis focused first on the \chandra\ data.  We
extracted the events in a large background region and used the CXC
tool {\tt psextract} to bin the source and background event data and
generate the appropriate response files.  We then fit the data using
the {\tt XSPEC} and {\tt Sherpa} packages independently.  As a caveat
to the results reported below, we note that the continuous-clocking
mode of ACIS has not yet been independently calibrated for spectral
purposes; our analysis depends on the calibration of the
timed-exposure ``Faint'' mode of ACIS S-3, which telemeters an
equivalent quantity of information about each event ($3\times 3$ pixel
islands).  To the extent that photon interaction times in the CCD
substrate are negligible compared to the CC single-row clocking time
of 2.85~ms, we expect this calibration to be accurate.  

Pure blackbody fits (with interstellar absorption) were unable to fit
the spectra, indicating that a harder spectral component was required.
Single-temperature thermal bremsstrahlung models were able to fit the
data but only with plasma temperatures so high that the resulting
spectra in the ACIS 0.2--10~keV range were little different from
simple power laws; thus a power-law component became our starting
point for the fits.

\begin{figure}
\plotone{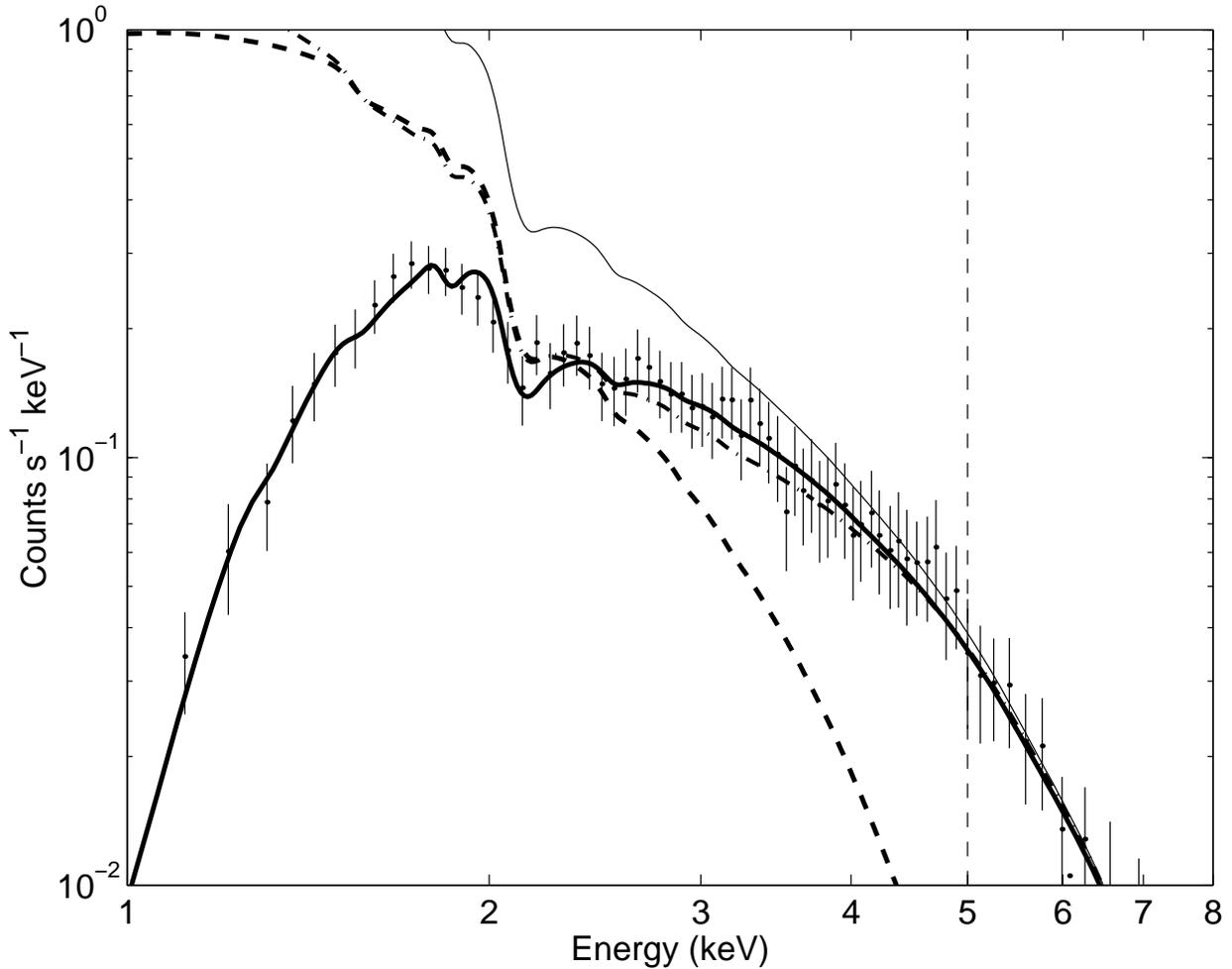}
\caption{\chandra\ spectrum of \sgr\ for the 2001~April~30 observation,
illustrating the contributions of various spectral components to the
final fit.  Plotted are: the data (points, with error bars); the
best-fit PL+BB model (thick line); its unabsorbed PL and BB components
(dashed and dash-dotted lines, respectively); and an unabsorbed
power-law fit to the 5--10~keV data only (thin solid line).  Above
5~keV (dashed vertical line) the unabsorbed power-law is
indistinguishable from the fully absorbed PL+BB model.  This indicates
that the region above 5~keV will provide the best measurement of the
power-law index for the fit.
\label{fig:sdecomp}}
\end{figure}

Single power-law (PL-only) models (with hydrogen absorption) were able
to fit the data satisfactorily (Table~\ref{tab:spec}).  However, the
resulting power-law indices are quite steep ($>$2.5), requiring high
column densities and implying relatively low high-energy fluxes,
especially when extrapolated to the \xte\ band.  Adding a blackbody
component to the models remedied these possible defects, but could not
be justified in a strict statistical sense ($F$-test probability of
62\% for the additional parameters in the joint fit).  To determine,
then, whether the PL-only fits were a realistic description of the
spectra, we examined the high-energy range of the data in greater
detail.  As shown in Fig.~\ref{fig:sdecomp}, above 5~keV the effects
of interstellar absorption, or of any $\approx$0.5~keV blackbody, on
the spectrum are negligible.

We therefore executed our power-law plus blackbody (PL+BB) fits in the
following manner. First, we fit a PL to the data in the 5--10~keV
range.  At these energies a BB component is superfluous, and fits are
insensitive to the precise amount of interstellar absorption.  As
indicated in Table~\ref{tab:spec}, the PL indices for these 5--10~keV
fits are significantly harder than those for the 1--10~keV PL-only
fits: the difference for the first observation is $2.7\sigma$ and for
the second $1.8\sigma$, giving an overall significance for the
distinction of 3.8$\sigma$.  We then froze the PL parameters at their
best fit 5--10~keV values, added a soft ($\sim 0.5$~keV) BB plus
interstellar absorption, and fit the full 1--10~keV dataset, allowing
the BB parameters as well as $N_{H}$ to vary -- note, however, that
$N_{H}$ was forced to be the same for both epochs.  This fit, with
fixed PL normalizations and power-law indices, required a soft (BB)
component at very high confidence.  After making this constrained fit,
we freed all parameters and fit a final time.  The results of this
final fit are given in Table~\ref{tab:spec}, along with the results of
the 1--10~keV and the 5--10~keV PL-only fits.

\begin{figure}
\plottwo{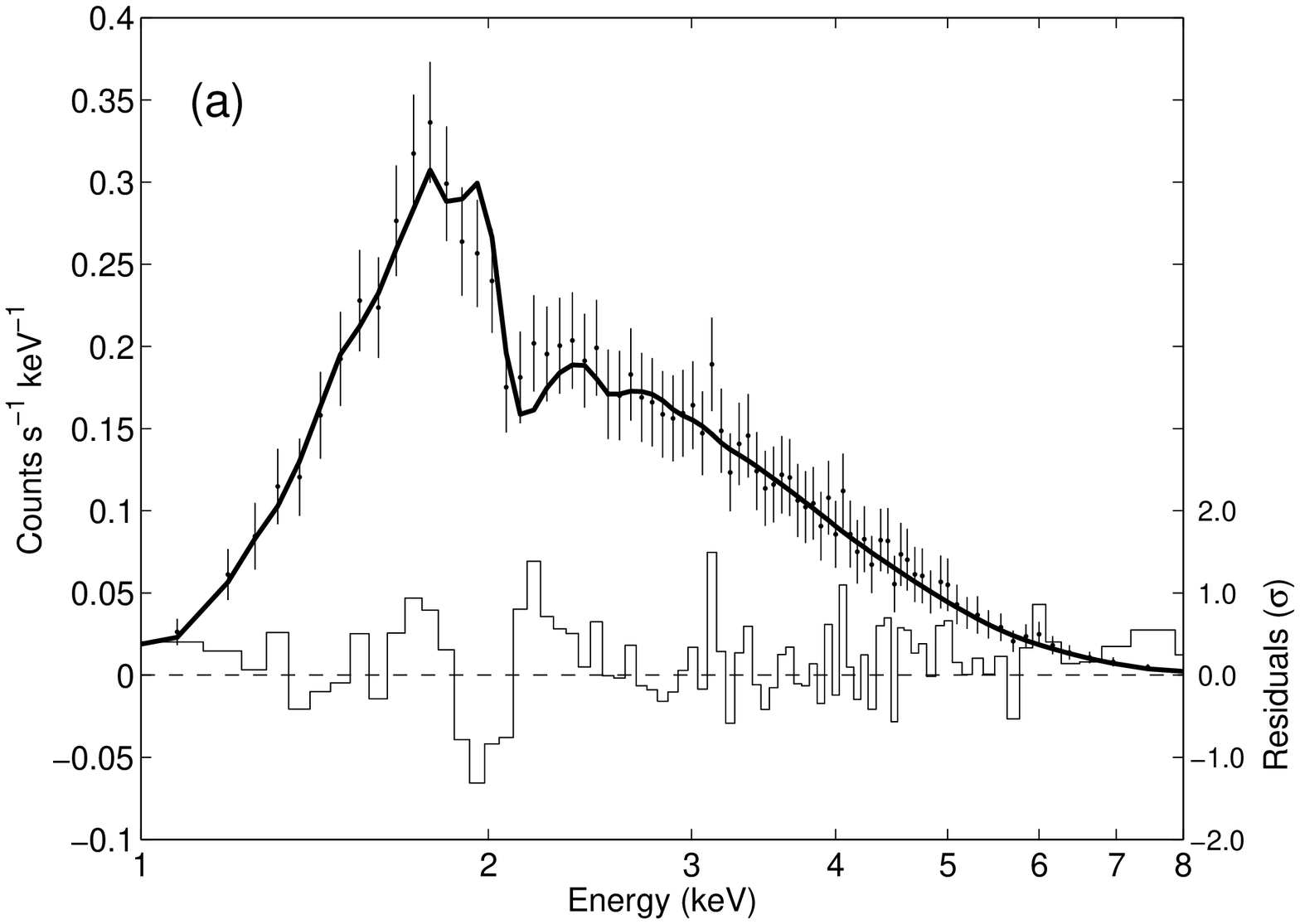}{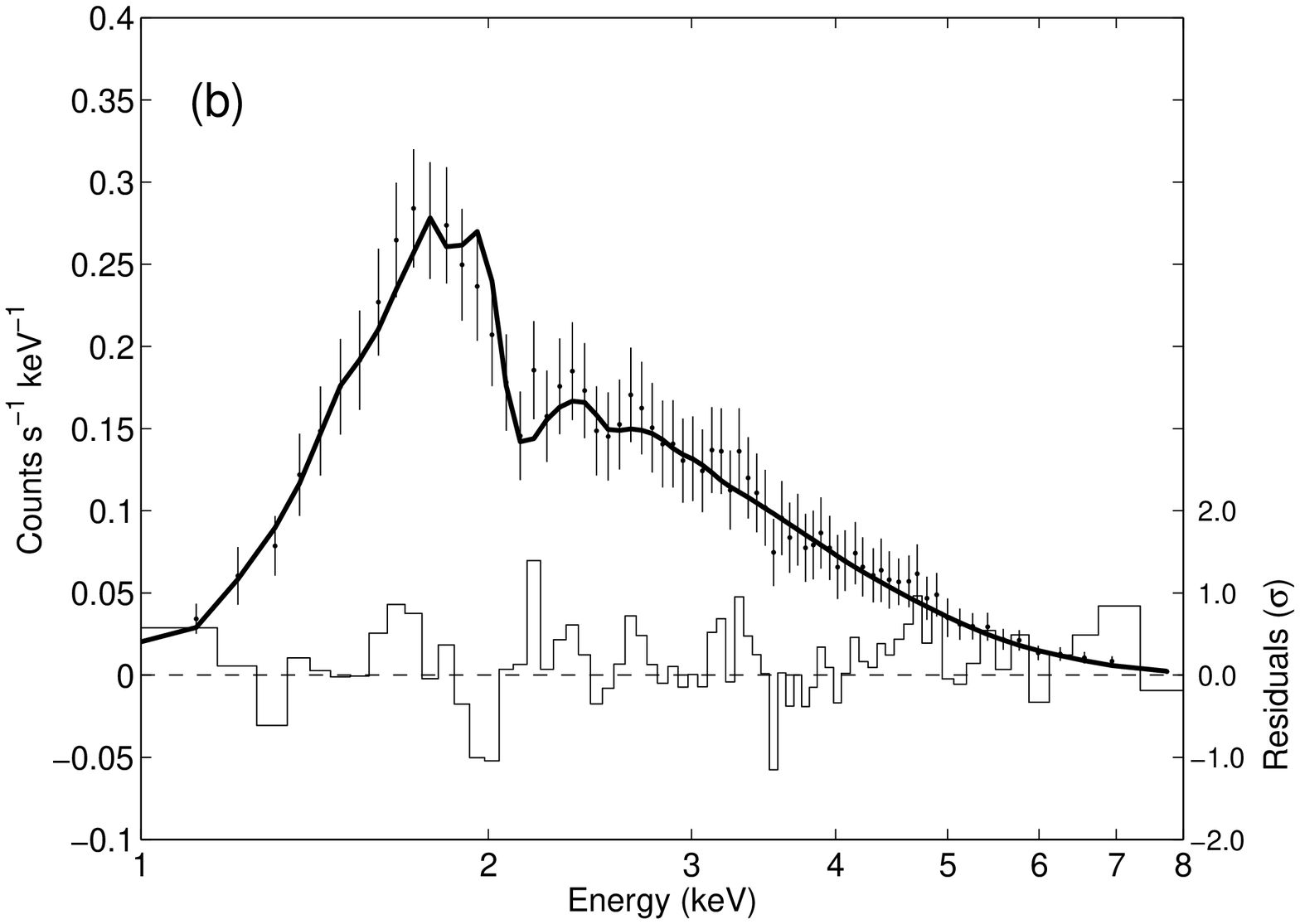}
\caption{\chandra\ spectra of \sgr\ for the (a) 2001~April~22 observation;
and (b) 2001~April~30 observation.  Plotted are data (points)
and best-fit PL+BB models (thick lines).  Residuals are plotted as the
thin lines below the spectra (right axis for scale).  The data have
been adaptively binned to have $\approx 60$~counts per bin.  The PL
fits are visually similar (see Table~\ref{tab:spec}).
\label{fig:spec}}
\end{figure}

Our best-fit PL+BB models are plotted in Figure~\ref{fig:spec}
(PL-only fits appear indistinguishable). We note that while \chandra\
does not have sufficient sensitivity at high energies ($>10$~keV) to
discriminate directly between the PL+BB and PL-only models, other
satellites, including \xte\ and \bepposax, do; see
Section~\ref{sec:phasespec} for our approach on this point.  Therefore
we also quote in Table~\ref{tab:spec} the 2--10~keV fluxes from the
models, as appropriate for comparison with other satellites.  While
the 0.5--10~keV fluxes are very similar for the PL or PL+BB models,
there are significant differences in the absorbed 2--10~keV fluxes of
the two models.

No narrow spectral features are apparent.  The absorption ``feature''
near 2~keV is likely to be an instrumental artifact (Si)\footnote{See
\texttt{http://asc.harvard.edu/cal/Links/Acis/acis/Cal\_prods/matrix/notes/Fl-esc.html}}.
The 2-$\sigma$ upper limit on the flux of a persistent 0.2-keV FWHM
emission line in the 5--7~keV range (c.f.\ \citealt{si00}) is
$10^{-5}\mbox{ photons cm}^{-2}\mbox{ s}^{-1}$, corresponding to an
equivalent width at 6.5~keV of less than 150~eV; narrower lines with
similar equivalent widths would have been readily apparent in the
data.

We attempted to extract time-averaged spectral information from the
\xte\ data as well.  However, with the faintness of the source (which
we estimate from our \chandra\ fits at 2~\ctsec\ for three active PCUs
of the \xte\ PCA), the high background ($\approx$100~\ctsec) -- some
of which is likely due to unresolved sources near the Galactic plane
-- and the non-imaging nature of the PCA, we have been unable so far
to obtain meaningful results.

\subsubsection{Phase-resolved spectral analysis}
\label{sec:phasespec}
\begin{figure}
\plottwo{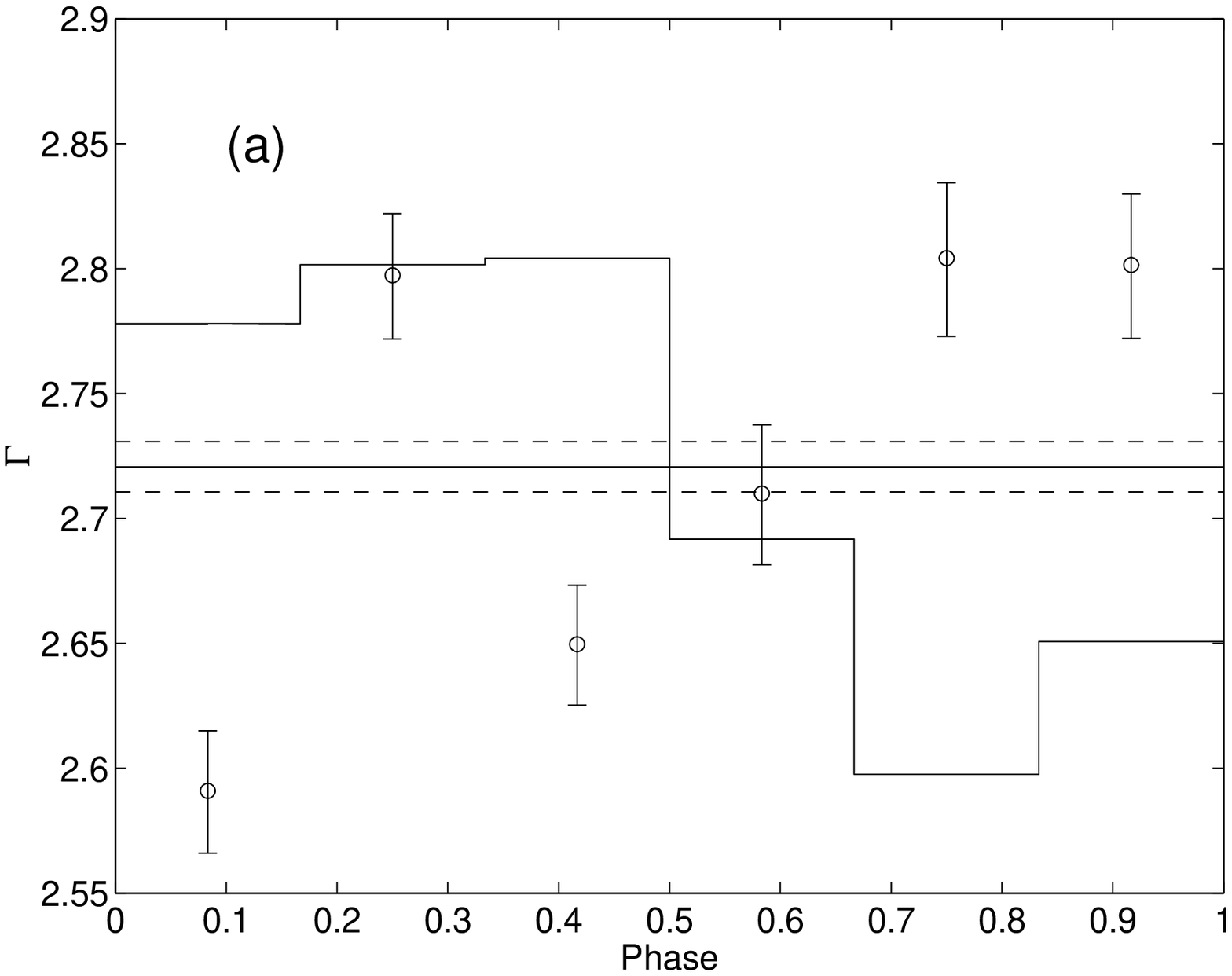}{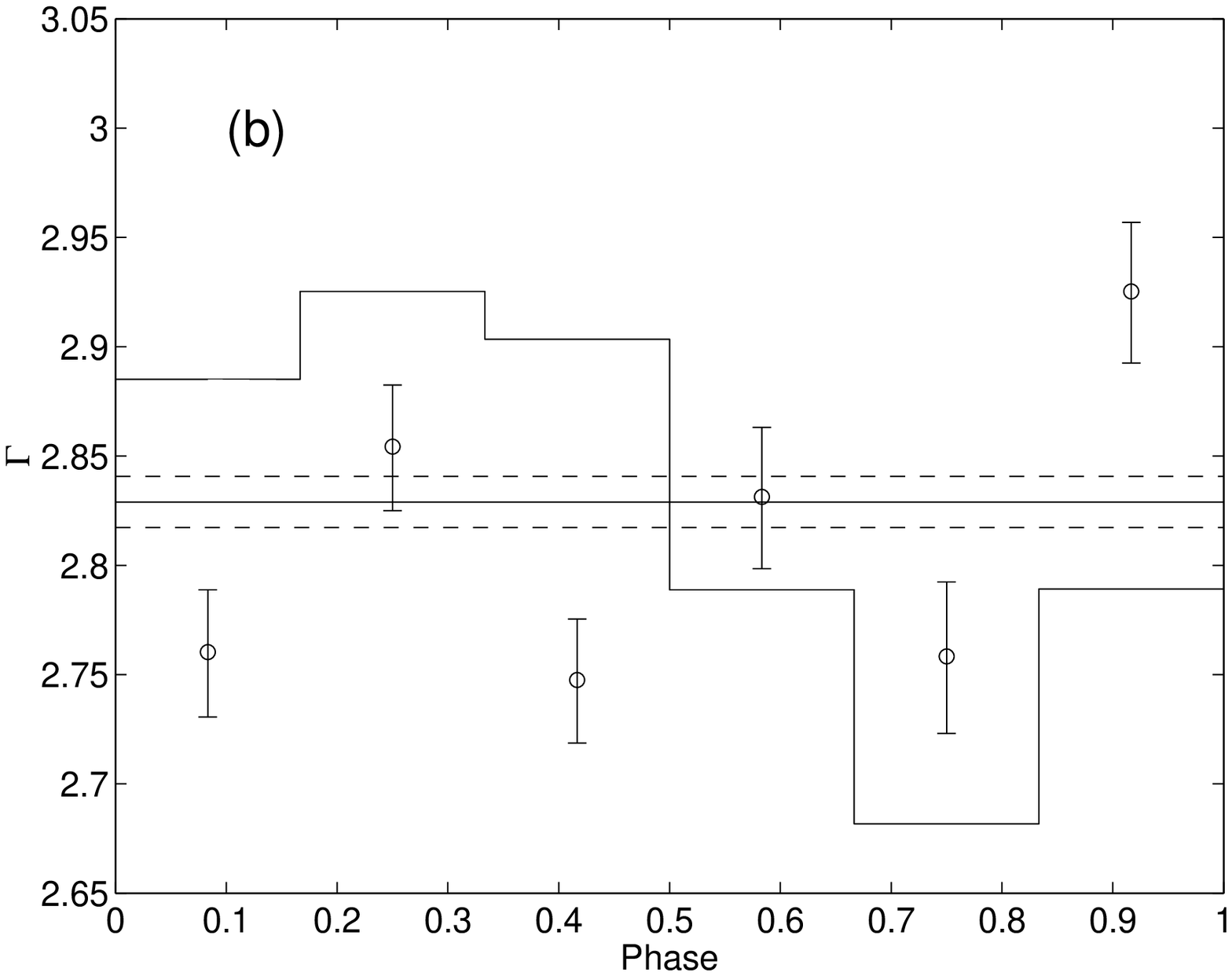}
\caption{Variation of the best-fit power-law index $\Gamma$ with pulse
phase for the (a) 2001~April~22 observation; and (b) 2001~April~30
observation.  The best-fit value of $N_{H} = 2.75\times
10^{22}$~\percm\ derived from the phase-averaged spectral analysis was
used for all of the fits (see Table~\ref{tab:spec}).  The
phase-average value of $\Gamma$ for each observation is plotted as the
line across the middle, with $\pm 1\sigma$ errors given by the dotted
lines.  The corresponding 0.3--10~keV pulse profiles are also
overplotted, with arbitrary scale.
\label{fig:ps}}
\end{figure}

To perform a phase-resolved spectral analysis we divided the \chandra\
events into six phase bins, according to the best fit period for each
observation.  We constructed a spectrum for each spectral bin
independently.  With the reduced counts of the phase-binned spectra we
were unable to discriminate between multicomponent spectral models and
fit a PL only, fixing $N_{H}$ to the best-fit value for the PL-only
fits to the phase-averaged data set ($2.75\times 10^{22}\mbox{
cm}^{-2}$; see Table~\ref{tab:spec}).  Whether or not this model is
accurate, the fits illustrate the gross variations in spectral shape
(hardness) with phase exhibited by the source.  We see in
Figure~\ref{fig:ps} that there are moderate variations across the
phase, with the beginning of the cycle harder than the end, and with
an additional softening at pulse-maximum.  The shape remains similar
over the two observations.

As mentioned previously, we were not able to make a direct comparison
with the \xte\ spectral results due to unresolved background emission
in the \xte\ data that corrupted the absolute flux levels.  However,
we were able to compare the fluxes for pulse ON$-$OFF.  Specifically,
we extracted spectral datasets for the $1/3$ of the phase around the
maximum (ON) and the minimum (OFF) of the pulse for both \chandra\ and
\xte\ for the second epoch of (overlapping) observations.  Our goal
was to use the two datasets in combination to make an independent test
of the reasonability of the PL+BB fits.

For the \xte\ data we fit the PCABACKEST-subtracted data in the
7--20~keV range for the ON and OFF datasets with a power-law plus
$\sim$7~keV Gaussian.  We are only interested in the flux difference
between these fits so the exact parameterization here is not crucial;
however, we do fix $N_{\rm H}$ in the fits to $2.0\times
10^{22}$~\percm, a reasonable value from the \chandra\ fits.  We
determined our uncertainties in the flux difference by exploring the
parameter space near the minimum for the ON and OFF datasets,
determining one-sigma flux errors for each, and combining these ON and
OFF errors in quadrature.  The \xte\ fits, combined with this
investigation of the errors, demonstrate an ON$-$OFF flux difference
of $4.0\pm 1.2 \times 10^{-12}$~\ergcms\ (7--20 keV).

When we fit the \chandra\ ON and OFF pulse data using a PL-only model,
the best-fit power-law photon indices for the two datasets are quite
similar, $\approx 2.8$, and the best-fit flux difference in the
extrapolated 7--20~keV band is $1.3\times 10^{-12}$~\ergcms.
Investigating the parameter space defined by the ON and OFF power-law
indices of the fit, which will have the largest effect on the
extrapolated \xte\ flux, we find that the maximum ON$-$OFF flux
allowed by the PL-only models is $1.57\times 10^{-12}$~\ergcms\
(3-$\sigma$ upper limit).  This value is 2$\sigma$ from the actual
\xte\ flux.  By contrast, the PL+BB fits give an ON$-$OFF flux
difference of $2.5\pm 1.4 \times 10^{12}$~\ergcms, which is only
0.8$\sigma$ different from the \xte\ value.

We feel that the combined weight of the 5--10~keV \chandra\ fits and
the 7--20~keV \xte-\chandra\ ON$-$OFF fits demonstrates that the BB
component of the spectral models is indeed required by the data.  We
note that \citet{ktw+01} reached the same conclusion by making fits to
the summed spectrum of the two \chandra\ observations.

\subsubsection{Bursts}
A cursory examination of the \xte\ data revealed several short
($<$0.25~s), intense ($>$30~\ctsecpcu) bursts -- as typical for SGRs
-- in each observation.  We therefore made a systematic search for
bursts in all data sets.  We constructed a 1/8-s-resolution light
curve for each observation and identified all bins with $>$4$\sigma$
fluctuations above background.  These bursts, detected in one or two
adjacent time-bins exclusively, are listed in Table~\ref{tab:bursts}.
Since we do not make a detailed investigation of the burst spectra,
burst fluences given in the table should be taken as suggestive only.

Exactly one burst is detected in the \chandra\ dataset.  This event
occurred at 2001~April~22.30, 9079.8~s after the start of the first
observation, and contains $\approx$14 photons over 0.27~s for a count
rate of $\approx 83$ times the quiescent rate.  This implies a
0.5--10~keV X-ray flux of $\approx \expnt{8}{-10}\mbox{ ergs
cm}^{-2}\mbox{ s}^{-1}$ (for either spectral model), and an unabsorbed
flux of $\expnt{1.9}{-9}\mbox{ ergs cm}^{-2}\mbox{ s}^{-1}$ (for the
PL+BB model) or $\expnt{3.6}{-9}\mbox{ ergs cm}^{-2}\mbox{ s}^{-1}$
(for the PL-only model).

All seven bursts detected during the second \xte\ observation occurred
during times of simultaneous observation by \chandra.  Examination of
the \chandra\ light curve reveals that several of these bursts were
accompanied by a mild excess in the ACIS count rate.  The excess
\chandra\ counts at the times of the \xte\ bursts provide an
independent demonstration of the consistency of the absolute timing
for both satellites, to a precision of $\lesssim$0.1~s.  In the context
of the \chandra\ continuous-clocking observation, this establishes the
position of the source on-chip, along the Y axis, to $\lesssim$35~pixels,
and the position of the source on-sky to $\lesssim$18~arcsec from its
August~1998 VLA position (the assumed location for all timing
analyses).  We make a more precise determination of the
two-dimensional source position below (Sec.~\ref{sub:spatial}).

\subsubsection{Pulsations}
Fourier power spectra of the two \chandra\ observations show a clear
pulse peak at the location of the known 5.17~s period of \sgr\
\citep{hlk+99}.  To make a more precise characterization of the pulse
period and phase at the start of each observation, we performed a
phase connection of the \chandra\ data; this also allowed us to test
for variability of the pulse strength over the course of the
observation.  The phase connection procedure was implemented by
dividing the observation into six sections and folding the data (a
barycenter-corrected light curve with 1/8-s resolution) in each
section about the pulse period.  From the folded pulse profile we
derived the phase at the start of each section, as well as a phase
uncertainty that we determined by Monte Carlo simulation.  The
connection of these phases then provided us with the value and
uncertainties for the observation's pulse period and starting pulse
phase.

We find the pulse profile of the first \chandra\ observation to be
largely sinusoidal.  There is evidence, at the 2.9-$\sigma$ level, for
power at the first harmonic of the pulse period (twice the main pulse
frequency); the harmonic power is $4.3^{+3.1}_{-1.8}$\% the power at
the fundamental if it is real.  Any power at the second harmonic is
less than 3.5\% the power of the fundamental at 90\%-confidence. 

\begin{figure}
\plotone{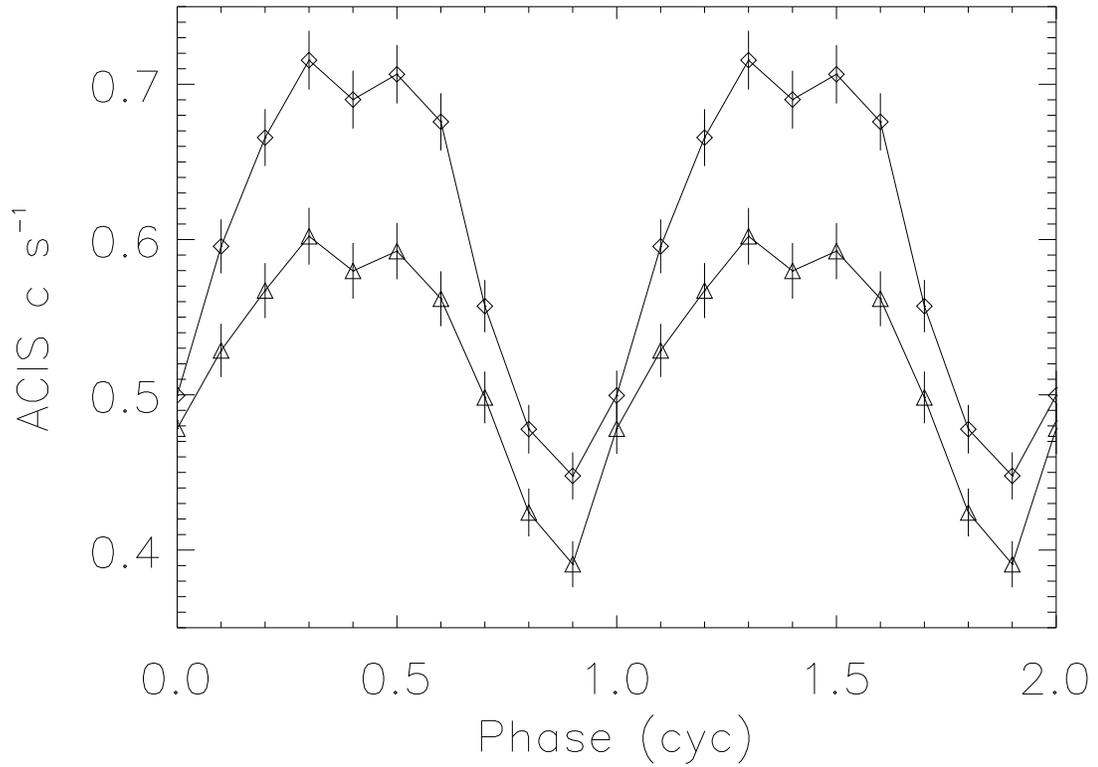}
\caption{Folded pulse profiles from the two \chandra\ observations
(first epoch: diamonds; second epoch: triangles; profiles are plotted
twice for clarity).  At the time of the second observation, \sgr\ was
exhibiting less flux, and, possibly, weaker modulation of the pulse
signal.  Both pulse profiles are highly sinusoidal, with $\sim$5\%
harmonic content.}
\label{fig:profiles}
\end{figure}

The pulse profile of the second \chandra\ observation has a first
harmonic detected with $>$3-$\sigma$ confidence.  Its power is
$10^{+7}_{-4}$\% the power at the fundamental, marginally consistent
with results from the first observation.  Any power at the second
harmonic is less than 7.2\% the power of the fundamental at
90\%-confidence for this observation.  Folded pulse profiles from the
two \chandra\ observations are shown in Figure~\ref{fig:profiles}.

We also performed a phase connection on the data from the \xte\
observations.  The procedure was identical; however, we used a
barycenter-corrected light curve of 1/8-s resolution divided into four
sections.  Uncertainties from this analysis were greater due to the
much larger background in the \xte\ data.

The results of our timing analysis are shown in Table~\ref{tab:pulse}.
A Bayesian period-estimator \citep{gl92} analysis of the data yielded
similar results.  The pulsed signal did not show detectable variation
in frequency or strength over the course of any observation,
consistent with prior and contemporaneous reports \citep{w+01} and
with expectations for a slow pulsar such as \sgr.  A global analysis
of our results yields a period derivative for \sgr\ of $\dot{P} =
9.2\pm 9.7 \times 10^{-11}$ at the epoch of the first \xte\
observation, MJD 52020.5, consistent with the more precise results of
\citet{w+01}.  There are no significant variations of the strength or
profile of the pulsations with energy for either \chandra\
observation.

\subsubsection{Localization}
\label{sub:spatial}
The one-dimensional spatial profile is consistent at close radii with
an unresolved source, having a Gaussian shape with $\sigma \approx
0\farcs3$ \citep[c.f.][]{mrf+01}.  \citet{ktw+01} have pointed out
that at larger radii ($>$5~arcsec) a scattering halo, the product of
X-ray scattering off of interstellar dust, is apparent.

\begin{figure}
\plotone{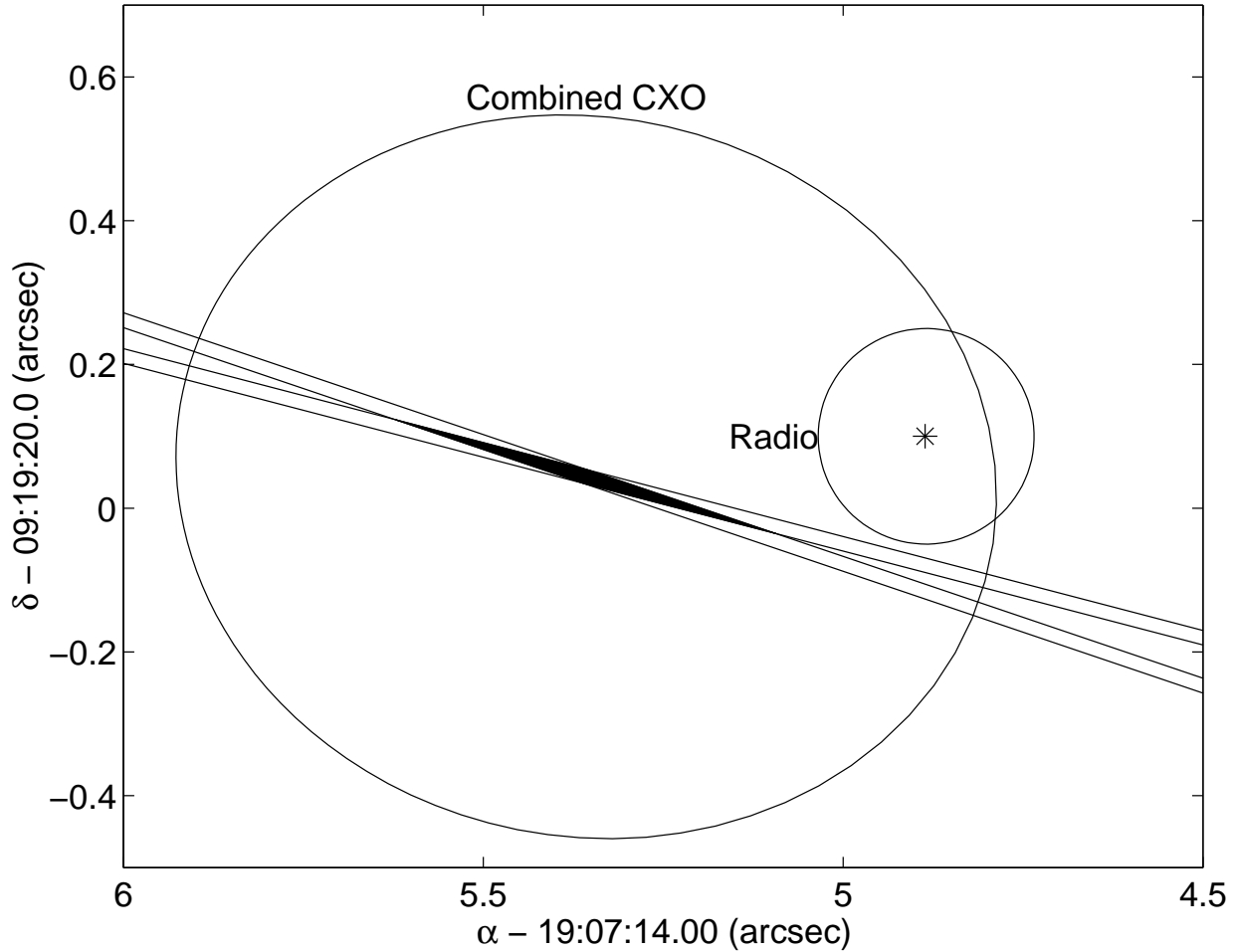}
\caption{Two-dimensional localization of \sgr.  The lines are the
localizations for each individual X-ray observation, with statistical
errors of $\sim 0\farcs02$ for each.  The shaded region is the
intersection of the two localizations, and represents the best-fit
position from the \chandra\ data.  The asterisk and circle marked
``Radio'' gives the position from \citet{fkb99} with its associated
$0\farcs15$ uncertainty.  The ellipse around the X-ray position
incorporates a $0\farcs5$ systematic error, identical for the two
observations, arising from the uncertain absolute aspect of \chandra;
this may be an underestimate.
\label{fig:pos}}
\end{figure}

With only one dimension of spatial information, an individual
observation does not give a precise two dimensional location (without
highly precise timing information), but merely confines the source to
a one-dimensional locus.  We were able, however, to use the two
observations together to obtain a two-dimensional localization as each
observation was taken at a slightly different orientation, with roll
angles differing by $4.8\degr$.  When the one-dimensional positions
are combined, there is a quasi-elliptical region of overlap centered
at $\alpha_{2000}=19^{\rm h}07^{\rm m}14\fs362$,
$\delta_{2000}=+09\degr19\arcmin20\farcs04$, with statistical
1-$\sigma$ error contours of semi-major axis $0\farcs28$, semi-minor
axis $0\farcs01$, and position angle $73\degr$ East of North
(Figure~\ref{fig:pos}).  This position has absolute systematic
uncertainties of up to $1\arcsec$ due to \chandra\ aspect errors, but
we have here assumed that the separate observations possess internally
consistent aspect determinations to $\lesssim$0.1~arcsec; this agrees
with our experience with other \chandra\ data sets.  Under this
assumption, the resulting localization is consistent to a high degree
of accuracy with the position of the August~1998 radio transient
\citep{fkb99}.

\section{Discussion}

\subsection{Radio}
\begin{figure}
\plotone{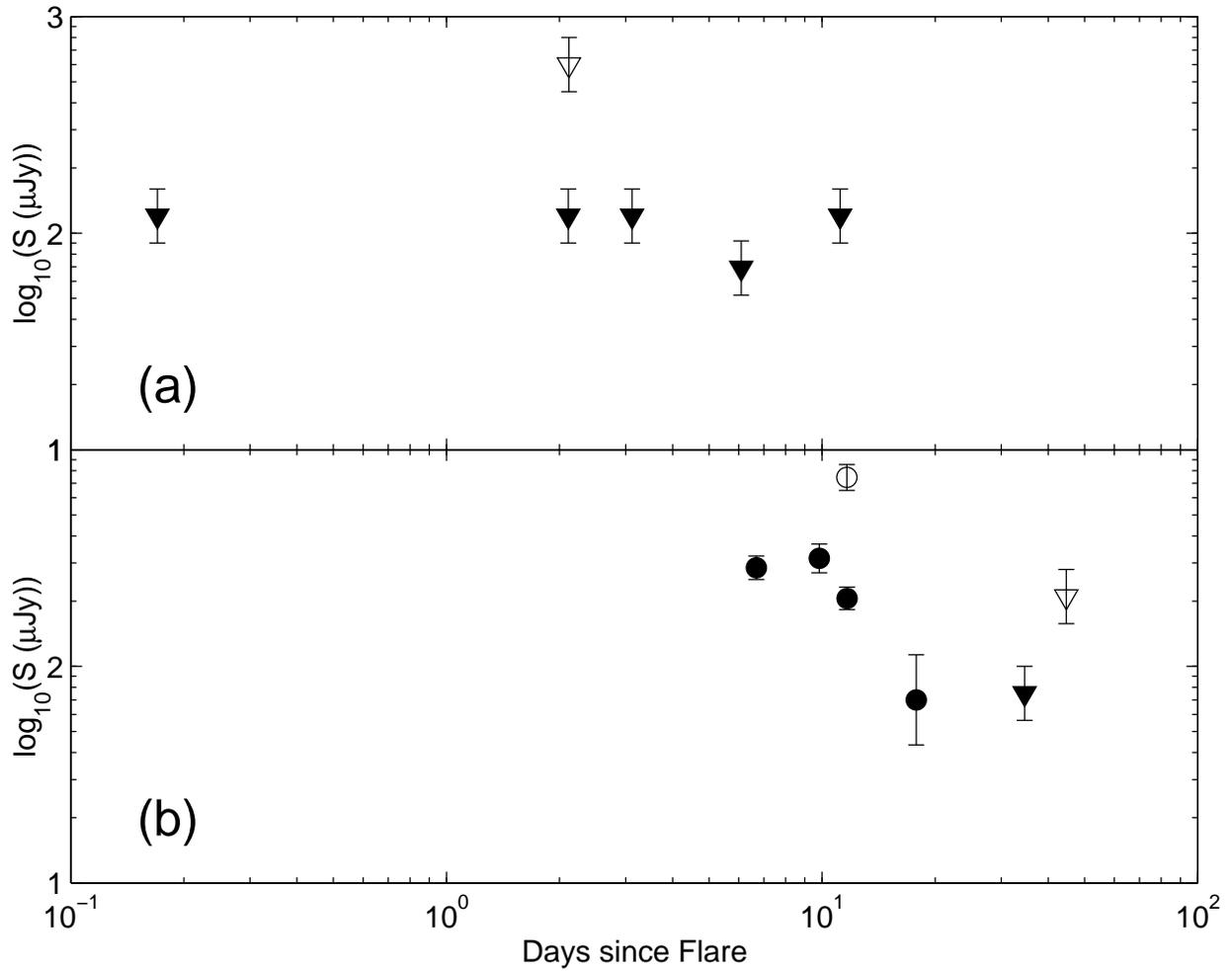}
\caption{VLA radio observations of \sgr\ in the aftermath of (a) the
current flare; and (b) the August~1998 flare \citep{fkb99}.
Detections are plotted as circles and upper limits as triangles;
corresponding radio frequencies are 1.4~GHz (open symbols) and 8.4~GHz
(filled symbols).  Upper limits are 3$\sigma$.
\label{fig:radio}}
\end{figure}

\citet{fkb99} reported, following the giant August~1998 flare, the
detection of a transient radio source. Their observations covered the
time interval from 1~week to 1~month after the burst. The source was
detected in the first observation, 1~week after the burst, and then
declined over the course of the following four observations (9--30~d;
Fig.~\ref{fig:radio}).  Thus -- at least for this giant flare -- the
radio source appears to have peaked about a week after the burst and
subsequently undergone a power-law decay.

For the April~2001 flare, we undertook VLA observations beginning
0.17~d after the event and ending almost two weeks later
(Table~\ref{tab:vla}). Despite our prompt radio observations, we did
not detect a radio source comparable in strength to the August~1998
flare at any of our five epochs of observation.

The fluence of the August~1998 flare was $10^{-2}$~\ergcm\ (here we
include the contribution to the fluence from the initial hard spike
and the subsequent softer afterglow; \citealt{fhd+01}).  In contrast,
the fluence of the April~2001 flare was $2.6\times 10^{-4}$~\ergcm\
\citep{h+01}.  The inferred peak flux of the transient radio source
for the August~1998 flare was about 400~$\mu$Jy in the 8.46-GHz band.
If the radio flux is proportional to the energy released by the flare
then we would expect a peak radio flux of 10~$\mu$Jy in the same band
some time $\sim$0.1--10~d after the current flare. In this context, as
seen from Table~\ref{tab:vla} and Fig.~\ref{fig:radio}, our failure to
detect a transient radio source is not surprising.

We end by noting a possibly interesting point. The origin of the
transient radio source from SGR outbursts is not well understood.  It
is clear that the radio emission is powered by the flare in some way.
Could it be that the radio emission arises from internal shocks of the
emitted particles?  Alternatively, the radio emission could be from
the shock of the ambient gas driven by the burst of particles (the
``afterglow'' model). In this context we note that the radio emission
of the August~1998 flare appears to peak, in the 8.46-GHz band, one
week after the burst, suggesting that the emission at this frequency
was optically thick.  If so, the flux at earlier epochs may well have
been higher at higher frequency.  One way to test this idea would be
to observe the burst source at higher frequencies ($>$20~GHz) at early
times.  Observatories currently capable of the requisite sensitivity
at these frequencies include the VLA, the Owens Valley Radio
Observatory, and the James Clerk Maxwell Telescope (SCUBA).

\subsection{X-ray}
With our X-ray observations, we are able to establish that the X-ray
source remains unresolved, in one dimension, even at \chandra's
extraordinary resolution.  Our upper limit of 0.6~arcsec on the FWHM
of any persistent extended emission translates into a physical extent
of less than $0.015 d_5$~pc, where $d_5$ is the distance to the source
divided by five kiloparsecs.  Recent suggestions that \sgr\ could be
as distant as 15~kpc \citep{vhl+00} may relax this constraint
somewhat, but in general the possibilities for localized plerionic
emission \citep[c.f.][]{bmg01} seem remote at this point.

Using the two \chandra\ observations in concert we are able to derive
a highly precise location for \sgr.  Although the uncertain \chandra\
absolute aspect probably implies an uncertainty of $\sim$1~arcsec in
this position, the best-fit location is less than 0.5~arcsec from the
location of the August~1998 radio flare \citep{fkb99}.  Our limit on
the proper motion of the source becomes relevant if one considers that
\sgr\ has been proposed to be associated with \snr, 12~arcmin distant,
and that the nominal spin-down age of \sgr\ was measured, at least
initially, to be just 700~yr \citep{ksh+99}.  Taken at face value this
would imply a proper motion of fully 1.0~arcsec per year, or
2.7~arcsec between August~1998 and May~2000, significantly exceeding
our limit.  The alternative possibilities -- that \sgr\ is not
associated with \snr, or that \sgr\ is significantly older than its
spin-down age would imply -- are consistent with our data.

Our X-ray spectral observations find the source initially in a state
of enhanced X-ray luminosity relative to the quiescent-state
observations of \citet{wkp+99a} and \citet{wkg+01}.  These past
observations found the source at an unabsorbed 2--10~keV flux of
$\approx 1\times 10^{-11}$~\ergcms\ on two occasions, 1997~May and
2000~March.  Our first observation finds the source at 20\% greater
flux, while the flux at the second observation is consistent with the
quiescent value (Table~\ref{tab:spec}).  The fading nature of this
excess emission is confirmed by reports of \bepposax\ observations at
earlier times \citep{mf+01}.  Spectral parameters for the thermal
component in the spectrum are consistent with values reported for the
source in quiescence \citep{wkp+99a,wkg+01,phh+01}, and despite the
20\% decline in total X-ray flux between our two observations, fitted
blackbody parameters show no significant variation -- indicating that
the variations in source flux are produced solely by the power-law
component in the spectrum.  Similar fading of the power-law component
alone was observed in the wake of the August~1998 giant flare of the
source \citep{wkp+99a}.  This stability of the blackbody component of
\sgr\ has been cited as evidence in favor of the magnetar model for
SGRs, with the blackbody produced by surface thermal emission and the
power law produced in the magnetosphere by, e.g., inverse Compton
effects \citep{wkp+99a}.  This interpretation must be further
strengthened by the observations reported here, as the stability of
the blackbody component of \sgr\ has now withstood two giant flares as
well as substantial excursions in X-ray flux.

The X-ray spectrum of \sgr\ thus provides an interesting link between
the SGR population and the related population of anomalous x-ray
pulsars (AXPs).  The AXPs (see \citealt{m99}) are a group of sources
that emit steady, pulsed X-ray emission ($L_{X} \sim 10^{35}\mbox{ erg
s}^{-1}$) with periods and spin-down rates similar to those of the
SGRs; indeed, \citet{td96} used spin properties to argue that the SGRs
and AXPs were related.  However, these spin properties are also shared
by a growing class of long-period radio pulsars \citep{kcl+99} which
seem otherwise unrelated to both the AXPs -- showing no persistent
bright X-ray emission \citep{pkc+00} -- and the SGRs -- showing no
bursting behavior.  We must therefore require more evidence than
similar spin properties to relate the SGRs and AXPs.

For some time the X-ray spectra of the two groups appeared to have
important differences: the SGRs had relatively hard power-law spectra
with photon indices $\Gamma \sim 2$ and negligible blackbody
contributions \citep{h99}, while the AXPs had softer spectra, with
$\Gamma \sim 4$ and $\sim 0.5$~keV blackbodies contributing up to 70\%
of the X-ray flux \citep{m99}.  But this situation has been changing.
Observations of the quiescent \sgrlmc\ found a photon index of 3.2 --
closer to the nominal AXP index than to the that of the other SGRs --
and possible evidence for a 0.5~keV blackbody \citep{k+00}.  This may
put the value of the photon index on a continuum related to burst
activity and magnetic field geometry for both groups.  Observations of
\sgr\ in quiescent and active states demonstrated the presence of an
underlying $\sim$0.5~keV blackbody \citep{wkp+99a,wkg+01}.  Updated
spectral fits of archival \asca\ data of SGRs and AXPs have shown that
both groups seem to possess blackbody components whose fraction of the
overall X-ray emission may constitute another unifying continuum
\citep{phh+01}.  Spectral fits to the AXP 1E~1048.1$-$5937 show that
it has a hard power-law component reminiscent of the SGRs
\citep{kgc+01a}.  And finally, optical and infrared observations of
SGRs \citep{k+00c,k+01} and AXPs \citep{hvkk00,hvkvk00} have shown
that the groups have similar X-ray-to-optical flux ratios, so that
this ratio may be a distinguishing characteristic of the two, as a
group \citep{hvkk00}.  All of these findings have strengthened
arguments for association between the AXPs and SGRs.

The blackbody component of the \sgr\ spectrum, with $kT_{\rm
BB}\approx 0.5$~keV and $R_{\rm BB}\approx 1.5 d_5$~km, has parameters
that are similar to those of other isolated NS candidates
\citep[c.f.][]{vbj+94}.  The relatively small emitting radius that we
find, significantly smaller than the nominal $\approx$10~km-radius NS,
is typically interpreted as either due to restricted emission from,
e.g., the NS polar caps, or as the result of temperature-dependent
opacity effects in the NS atmosphere \citep{rbb+99,phh+01}.  The
latter scenario would allow for closer distances, lower surface
temperatures, and, potentially, emission from the entire NS surface
\citep{phh+01}.

The absence of any narrow spectral features, to equivalent widths of
less than 150~eV, is somewhat surprising given the detection by
\citet{si00} of a strong, 400-eV equivalent width, $\approx$6.4-keV
emission line in the \rxte\ spectrum of a 1998 August~29 burst of
\sgr.  \citet{si00} discuss two possible interpretations for the
feature they observe: first, that it may result from fluorescence of
relatively cool iron in the near vicinity of the NS; and second, that
it may result from proton or alpha particle (He$^4$) cyclotron
transitions in the SGR magnetosphere; these ions would have been
liberated by the closely-preceding giant flare of 1998 August~27.
However, if the line resulted from iron fluorescence then we would
expect, with \chandra's superior spectral resolution, to have
substantially greater sensitivity with the current observations; note
that the original emission feature was unresolved by the \rxte\ PCA.
Similarly, if the line resulted from cyclotron emission by He$^4$ or H
ions liberated by the preceding giant flare, then we would expect to
observe such cyclotron emission in the wake of the current 2001
April~18 flare as well.  Both of these scenarios would generate line
emission in bursting and persistent source spectra alike.  Our upper
limits on the presence of any such feature in the persistent source
spectrum therefore seem to imply that the line emission is a function
of the properties of (at least some of) the bursts alone.

\section{Conclusions}
We have observed the soft-gamma repeater \sgr\ with high-resolution
X-ray (\chandra, \xte) and radio (VLA) observatories in the immediate
aftermath of its 2001~April~18 giant flare.  Our detailed study of the
\chandra\ and \xte\ X-ray spectra reveals the presence of an
underlying thermal component in the spectrum.  This thermal component,
which we model as a blackbody, has an effective temperature of
$kT_{\rm BB}\approx$0.5~keV and an effective blackbody radius of
$R_{\rm BB}\approx 1.5 d_5$~km, where $d_5$ is the distance to the
source divided by 5~kpc.  This two-component PL+BB spectrum is
strongly reminiscent of the spectra of the anomalous X-ray pulsars,
further strengthening the association between these two intriguing
classes of neutron star.

We detect enhanced, fading X-ray emission from the source, which is
modulated by the source's known 5.17~s pulsations and, intermittently,
by brief bursts of hard X-rays.  Detailed studies of the pulsations
will be able to determine whether any ``glitch'' of the source was
associated with the April~18 flare.  We are able to localize the
source to $\lesssim$1~arcsec, and find that the current position remains
consistent with the \citet{fkb99} radio position from August~1998.
This corresponds to an upper limit of $\lesssim$1~arcsec per year on any
proper motion of \sgr.

The upper limits we derive on any radio emission from \sgr\ in the
0.2--11~d following the flare may have implications for physical
models of the post-flare radio emission.  Higher-frequency ($>$20~GHz)
observations in the immediate aftermath of future flares will provide
stronger constraints for models that predict optically thick emission
at lower frequencies.

%%%%%%%%%%%%%%%%%%%%%%%%%%%%%%%%%%%%%%%%%%%%%%%%%%

\acknowledgements

The authors would like to thank Harvey Tananbaum, Jean Swank, and the
\chandra\ and \rxte\ operations teams for their rapid and effective
response to this target of opportunity.  We would also like to thank
G.~Taylor, C.~Carilli, T.~Soifer, and J.~Condon for donations of VLA
time. DLK is supported by the Fannie \& John Hertz Foundation.

%%%%%%%%%%%%%%%%%%%%%%%%%%%%%%%%%%%%%%%%%%%%%%%%%%

%\bibliographystyle{apj}
%\bibliography{myrefs,magrefs,casA,xray}

%%%%%%%%%%%%%%%%%%%%%%%%%%%%%%%%%%%%%%%%%%%%%%%%%%
%%%%%%%%%%%%%%%%%%%% TABLES %%%%%%%%%%%%%%%%%%%%%%
%%%%%%%%%%%%%%%%%%%%%%%%%%%%%%%%%%%%%%%%%%%%%%%%%%

\begin{deluxetable}{c c c c c c c}
\tablecaption{Summary of 0.5--10~keV spectral fits to \chandra\ data.\label{tab:spec}}
\tablewidth{36pc}
\tablehead{
\colhead{Parameter} & \multicolumn{2}{c}{PL} 
& \multicolumn{2}{c}{5--10~keV PL} & \multicolumn{2}{c}{PL+BB} \\
& \colhead{Obs 1} & \colhead{Obs 2} & \colhead{Obs 1} & \colhead{Obs 2} &
\colhead{Obs 1} & \colhead{Obs 2}}
\startdata
$N_{H}$ ($10^{22}\mbox{ cm}^{-2}$)\tablenotemark{a} & 
\mc{2}{c}{$2.75(5)$} & \mc{2}{c}{\nodata}& \mc{2}{c}{2.3(1)} \\
\tableline
PL Index $\Gamma$ & $2.72(5)$ & $2.82(5)$ & $1.9(3)$ & $2.1(4)$ &
2.0(2) & 1.9(3) \\
PL Norm\tablenotemark{b} & $0.0131(8)$ & $0.0126(8)$ & $0.003(1)$ &
$0.003(2)$ & 0.004(1) & 0.003(1) \\
PL Flux\tablenotemark{c} & 9.2 & 7.6  & \nodata & \nodata & 7.9 & 6.3\\
\tableline  
BB $kT$ (keV) & \nodata & \nodata & \nodata & \nodata & 0.52(3) & 0.49(3)\\
BB Norm\tablenotemark{d} & 
  \nodata & \nodata & \nodata & \nodata & 0.008(2) & 0.011(3)\\
BB Flux\tablenotemark{c} & \nodata & \nodata & \nodata & \nodata & 1.8
  & 1.8 \\
\tableline 
Total Flux\tablenotemark{c} & 9.2 & 7.6 & \nodata & \nodata & 9.7 & 8.1\\
Unabs.\ Flux\tablenotemark{c} & 42.6 & 39.5 & \nodata & \nodata &
  23.1 & 19.5 \\
2--10~keV Flux\tablenotemark{c} & 8.5 & 6.9 & \nodata &
  \nodata & 9.0 & 7.5 \\
Unabs.\ 2--10~keV Flux\tablenotemark{c} & 11.7 & 9.7 & \nodata &
  \nodata & 11.6 & 9.6 \\
\tableline
$\chi^{2}$ & 279.8 & 227.5 & 32.8 & 25.2 & 276.2 & 220.7\\
DOF & 318 & 287 & 59 & 44 & 318 & 288 \\
$\chi^{2}/{\rm DOF}$ & 0.88 & 0.84 & 0.56 & 0.57 & 0.87 & 0.77 \\
Total $\chi^{2}/{\rm DOF}$\tablenotemark{e} & \mc{2}{c}{0.84} &
  \mc{2}{c}{0.56} & \mc{2}{c}{0.82}\\ 
\enddata
\tablenotetext{a}{Held constant over the two observations for a given model.}
\tablenotetext{b}{In units of $\mbox{photons cm}^{-2}\mbox{
s}^{-1}\mbox{ keV}^{-1}$
at 1~keV.}
\tablenotetext{c}{All fluxes are 0.5--10~keV, absorbed, unless otherwise
specified; flux units are $10^{-12}\mbox{ erg cm}^{-2}\mbox{ s}^{-1}$.}
\tablenotetext{d}{In units of $0.44 (R_{10}/d_{5})^{2}$, where
$R=10R_{10}\mbox{ km}$ is the source radius.}
\tablenotetext{e}{Total reduced $\chi^{2}$ for a model, incorporating
both observations.}
\tablecomments{Number in parentheses is 1-$\sigma$ uncertainty on last digit.}
\end{deluxetable}

\begin{deluxetable}{l r r r}
\tablewidth{25pc}
\tablecaption{Observed Bursts \label{tab:bursts}}
\tablehead{
\colhead{~} &
\colhead{Time} & 
\colhead{Counts} & 
\colhead{Fluence} \\
\colhead{Date / Inst.} &
\colhead{(TDB)} & 
\colhead{(raw)} & 
\colhead{($\times 10^{-11}$~cgs)}
}
\startdata
21~April \xte\ (3) & 08:05:41.38 &   20  &   6.3 \\
                   & 08:08:56.88 &   31  &   9.9 \\
                   & 08:34:41.88 &   30  &   9.6 \\
                   & 08:34:58.51 &   38  &  12.2 \\
                   & 08:38:45.25 &  166  &  52.7 \\
                   & 08:42:18.13 &   27  &   8.6 \\
                   & 08:42:19.01 &   80  &  25.3 \\
                   & 08:45:50.26 &   65  &  20.7 \\
21~April \xte\ (4) & 09:49:38.48 &   17  &   4.0 \\
                   & 09:50:26.86 &   18  &   4.2 \\
                   & 10:17:23.61 &   31  &   7.5 \\
                   & 10:22:26.23 &   22  &   5.2 \\
21~April \xte\ (2) & 11:19:11.83 &  150  &  71.3 \\
22~April \chandra  & 07:18:00.30 &   10  &  36.5 \\
1~May \xte\ (4)    & 01:49:06.89 &   65  &  15.4 \\
                   & 02:16:55.26 &   28  &   6.7 \\
                   & 02:35:08.01 &   16  &   3.8 \\
                   & 02:35:43.39 &   19  &   4.4 \\
                   & 02:36:56.64 &   15  &   3.5 \\
                   & 02:37:28.51 &   18  &   4.3 \\
                   & 02:37:32.64 &   16  &   3.9 \\
1~May \xte\ (3)    & 03:36:40.97 &   21  &   6.7 \\
                   & 04:01:26.22 &  113  &  35.9 \\
\enddata
\tablecomments{Numbers in parentheses for \xte\ observations indicate
the number of active PCUs at the time of observation.  Times are
corrected to the solar-system barycenter (TDB).  Estimated fluences in
units of $10^{-11}$~\ergcm\ are calculated using the following
conversion factors: 1~\xte~c~PCU$^{-1}$ $\approx$ $9.5\times
10^{-12}$~\ergcm\ (7--20 keV); 1~\chandra~ACIS~c $\approx$ $3.7\times
10^{11}$~\ergcm\ (0.1--11 keV).  }
\end{deluxetable}

\begin{deluxetable}{l c c l c}
\tablewidth{30pc}
\tablecaption{Pulse Timing Analysis\label{tab:pulse}}
\tablehead{
\colhead{~} & \colhead{~} & \multicolumn{3}{c}{{Pulsations}} \\
\colhead{Mission} &
\colhead{Epoch (MJD)} & 
\colhead{Phase (cyc.)} &
\colhead{Period (s.)} & \colhead{Strength (\%rms)}
}
\startdata
\xte     & 52020.5~ & 0.76(9) & 5.17274(22)  & \nodata  \\
\chandra & 52021.25 & 0.84(2) & 5.172908(40) & 16.2(9)~ \\
\chandra & 52030.0~ & 0.46(2) & 5.172947(65) & 13.4(11) \\
\xte     & 52030.0~ & 0.62(9) & 5.17321(21)  & \nodata  \\
\enddata
\tablecomments{Times are corrected to the solar-system barycenter
(TDB); MJD is ${\rm JD}-2,400,000.5$.  Phases reported refer to the
phase of the sine wave at the fundamental frequency.  Uncertainties in
the last significant digit(s) are shown in parentheses.  RMS pulse
strengths have been corrected for background and include the
contributions of the fundamental and first harmonic power.}
\end{deluxetable}

\begin{deluxetable}{c c c c c}
\tablewidth{30pc}
\tablecaption{Summary of VLA Observations\label{tab:vla}}
\tablehead{
\colhead{Date of} & \colhead{Days after} & \colhead{Frequency} &
\colhead{Duration} & \colhead{rms} \\
\colhead{Observation (UT)} & \colhead{Flare} & \colhead{(GHz)} &
\colhead{(min)} & \colhead{($\uJy$)}
}
\startdata
2001~Apr~18.58 & 0.17 & 8.4 & 15 & 40 \\
2001~Apr~20.52 & 2.11 & 8.4 & 15 & 40 \\
2001~Apr~20.53 & 2.12 & 1.4 & 15 & 200 \\
2001~Apr~21.53 & 3.12 & 8.4 & 15 & 40 \\
2001~Apr~24.50 & 6.09 & 8.4 & 60 & 20 \\
2001~Apr~29.58 & 11.2 & 8.4 & 15 & 40 \\
\enddata
\tablecomments{All observations were in B configuration.}
\end{deluxetable}

%%%%%%%%%%%%%%%%%%%%%%%%%%%%%%%%%%%%%%%%%%%%%%%%%%
%%%%%%%%%%%%%%%%%%%%% FIGURES %%%%%%%%%%%%%%%%%%%%
%%%%%%%%%%%%%%%%%%%%%%%%%%%%%%%%%%%%%%%%%%%%%%%%%%

\end{document}